\documentclass[a4paper, 11pt]{article}
\date{December 2005}
\usepackage{epsfig,,booktabs}
\newcommand{\be}{\begin{equation}}
\newcommand{\ee}{\end{equation}}
\newcommand{\ba}{\begin{eqnarray}}
\newcommand{\ea}{\end{eqnarray}}
\newcommand{\bes}{\begin{eqnarray}}
\newcommand{\ees}{\end{eqnarray}}
\newcommand{\bi}{\begin{itemize}}
\newcommand{\ei}{\end{itemize}}

\newcommand{\dg}{^\dagger}

\newcommand{\Nf}{N_{\rm f}}

\newcommand{\Lmsbar}{\mathop{\Lambda_{\overline{\rm MS}}}}
\newcommand{\Lmsbartwo}{\mathop{\Lambda^{(2)}_{\overline{\rm MS}}}}

\newcommand{\Fk}{\mathop{F_{\rm K}}}

\newcommand{\Fps}{\mathop{F_{\rm PS}}}
\newcommand{\Mps}{\mathop{M_{\rm PS}}}
\newcommand{\Mv}{\mathop{M_{\rm V}}}
\newcommand{\Gps}{\mathop{G_{\rm PS}}}
\newcommand{\fP}{\mathop{f_{\rm P}}}
\newcommand{\fA}{\mathop{f_{\rm A}}}
\newcommand{\fAimpr}{\mathop{f_{\rm A,I}}}
\newcommand{\ca}{\mathop{c_{\rm A}}}
\newcommand{\meffa}{\mathop{m_{\rm eff}^{\rm A}}}
\newcommand{\meffp}{\mathop{m_{\rm eff}^{\rm P}}}
\newcommand{\meffv}{\mathop{m_{\rm eff}^{\rm V}}}
\newcommand{\feff}{\mathop{F_{\rm eff}}}
\newcommand{\geff}{\mathop{G_{\rm eff}}}
\newcommand{\fone}{\mathop{f_{1}}}
\newcommand{\kV}{\mathop{k_{\rm V}}}

\newcommand{\mq}{\mathop{m_{\rm q}}}

\newcommand{\tauint}{\mathop{\tau_{\rm int}}}

\newcommand{\half}{{\textstyle\frac{1}{2}}}

\newcommand{\<}{\langle}
\renewcommand{\>}{\rangle}
\newcommand{\eq}[1]{Eq.~(\ref{#1})}

\newcommand{\fig}{Fig.~}
\newcommand{\tab}[1]{Tab.~\ref{#1}}
\newcommand{\la}{\label}

\newcommand{\mbar}{{\bar m}}

\newcommand{\hQ}{\hat Q}
\newcommand{\hQa}{\hat Q_A}
\newcommand{\Mee}{M_{\rm ee}}
\newcommand{\Meo}{M_{\rm eo}}
\newcommand{\Moe}{M_{\rm oe}}
\newcommand{\Moo}{M_{\rm oo}}
\newcommand{\Mmee}{M^{-1}_{\rm ee}}
\newcommand{\Mmoo}{M^{-1}_{\rm oo}}
\newcommand{\gf}{\gamma_5}

\def\zp{Z_{\rm P}}
\def\za{Z_{\rm A}}
\def\bA{b_{\rm A}}
\def\bP{b_{\rm p}}
\def\gbar{\bar{g}}

\def\fm{{\rm fm}}

\begin{document}
\begin{titlepage}
\begin{flushleft}
CERN-PH-TH-2008-079\\
DESY 08-043\\
HU-EP-08/12 \\
MIT-CTP 3942\\
MS-TP-08-5\\
SFB/CPP-08-22 
\end{flushleft}
\begin{centering}
\vfill

\centerline{\bf \Large  Scaling test of two-flavor O($a$)--improved lattice QCD}

\vspace{1.5cm}
\centerline{\psfig{file=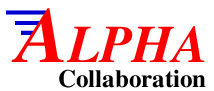,angle=0,width=3cm}}
\vspace{1.0cm}

{\bf  Michele~Della~Morte$^{a,f}$,
Patrick~Fritzsch$^{b}$,
Harvey~Meyer$^{c,e}$,
Hubert~Simma$^{d,e}$, Rainer~Sommer$^{e}$,
Shinji~Takeda$^{f}$,\\ Oliver~Witzel$^{f}$, Ulli~Wolff$^{f}$}
\vspace{0.55cm}

\emph{$^a$ CERN, Physics Department, TH Unit, CH-1211 Geneva 23, Switzerland\\
$^b$ Institut f\"ur Theoretische Physik, Westf\"alische Wilhelms-Universit\"at M\"unster,\\
 Wilhelm-Klemm-Strasse 9, D-48149 M\"unster, Germany\\
$^c$  Center for Theoretical Physics, Massachusetts Institute of Technology, \\
Cambridge, MA 02139, USA\\
$^d$ Universit\`a di Milano Bicocca, Piazza della Scienza 3, 20126 Milano, Italy\\
$^e$ DESY, Platanenallee 6, 15738 Zeuthen, Germany\\
$^f$ Institut f\"ur Physik, Humboldt Universit\"at, Newtonstr. 15, 12489 Berlin, Germany}\\
\vspace*{2.0cm}

\end{centering}
\centerline{\bf Abstract}
\vspace{0.1cm}
\noindent
We report on a scaling test of several mesonic
observables in the non-perturbatively O($a$) improved Wilson theory
with two flavors of dynamical quarks. The observables are constructed
in a fixed volume of $2.4\,\fm \times (1.8\,\fm)^3$ with 
Schr\"odinger functional boundary conditions. 
No significant scaling violations are found. Using the kaon mass determined
in \cite{cernI}, we update our estimate of the Lambda parameter
to $\Lmsbartwo/m_{\rm K} = 0.52(6)$.

\vfill
\end{titlepage}

\setcounter{footnote}{0}
\section{Introduction\la{sec:intro}}

In this article we summarize the results of a set of simulations
of QCD with two degenerate flavors of quarks employing Schr\"odinger functional
boundary conditions~\cite{Luscher:1992an}. The range of quark masses
covered corresponds to a ratio of the pseudoscalar mass to the vector mass,
$\Mps/\Mv$, in the interval $[0.4,0.75]$. 
Our final goal is to compute the fundamental parameters of perturbative QCD, 
namely 
the scale parameter $\Lambda$ and the quark masses $M_{\rm q}$,
in units of a hadronic observable such as the Kaon decay constant $\Fk$.
We emphasize our effort to control all systematics. Here we focus on 
cutoff effects and reach (for one quark mass) a lattice spacing
that is smaller than those previously achieved in large-volume simulations
of the O($a$) improved Wilson action~\cite{cernI,cernII,Brommel:2006ww,Aoki:2002uc}.

While simulations of QCD with at least $\Nf=2+1$ flavors of sea quarks
are mandatory to provide accurate 
non-perturbative predictions with direct phenomenological implications,
in our view the $\Nf=2$ theory represents a framework well suited to address 
a number of fundamental aspects of low-energy QCD that have not been 
clarified yet, a couple of which we shall presently mention.

One such question is the $\Nf$ dependence of 
$\Lmsbar/\Fk$ and  $M_{\rm s}/\Fk$.
Since these quantities have been computed in the 
quenched theory~\cite{Capitani:1998mq,Garden:1999fg},
it is interesting to know the separate effects of 
the (up, down) quarks and those of the strange quark.
To our knowledge, the influence of the strange sea quarks
on hadronic observables has not been demonstrated very clearly so far.

Secondly, it is important to determine the quark mass at which one-loop 
SU(2) chiral perturbation theory becomes accurate at the (say) $3\%$ level.
We see a strong motivation to address this question in the $\Nf=2$
theory, with one parameter less to tune on the QCD side.
And with a small number of low-energy constants in the 
chiral perturbation theory, this is probably the cleanest way
to establish the latter as the low-energy description of QCD from first principles.
Given the level of accuracy one is interested in, all 
sources of systematic error have to be addressed. In particular any observed 
non-linearity in the quark-mass dependence of $\Fps$ and $M_{\rm PS}^2$ 
must first be shown to survive the infinite volume limit before 
it can be claimed that the chiral logarithms have been observed.
Cutoff effects represent an additional source of systematic uncertainty,
which is computationally expensive to reduce. In particular,
cutoff effects may be larger
in the presence of sea-quarks~\cite{lat03:rainer}. With Wilson fermions, 
even in their O($a$) improved version that we employ,
it is well known that the chiral limit does not commute with 
the continuum limit, implying that at fixed lattice spacing $a$
cutoff effects become large below some quark mass.
It is therefore important to control cutoff effects
as one proceeds to simulate deeper in the chiral regime.

In the quenched work~\cite{guagnelli}, rather accurate results were obtained in
the pseudoscalar and vector channels using the Schr\"odinger functional.
In this paper we carry over this computational setup to the $\Nf=2$ theory.
The accuracy achieved~\cite{guagnelli} on masses 
was comparable to the calculations performed
with periodic boundary conditions, and for decay constants
the Schr\"odinger functional even proved to be the superior method.
This is different when dynamical fermions are present.  
As shown in \cite{lat07:rainer}
multi-pion excited states contribute significantly.
For a computation of ground state masses and matrix elements
they have to be supressed by a rather large time extent of the 
Schr\"odinger functional -- in particular when the quark mass is low. 
In this 
situation it is more practical to employ (anti)periodic boundary conditions 
with the associated translation invariance in time. 
We can nonetheless use our simulation results to perform a first 
scaling test of the $\Nf=2$ O($a$)-improved theory at low energies. 
Note that at high energies and correspondingly small lattice spacings 
excellent scaling has been seen 
\cite{DellaMorte:2005kg,alpha2}. Besides the scaling test we give some details
of our simulations including the algorithmic performance (section 2).

\section{Lattice simulations}
Our discretization consists of the  Wilson gauge action
and the non-perturbatively O($a$) improved 
Wilson quark action, with $c_{\rm sw}$ given in~\cite{csw}.
The algorithm and solver used in the present simulations
have been described in some detail in~\cite{Meyer:2006hx,DellaMorte:2003jj}.
Using the notation of~\cite{jliu} for the hopping terms of 
the Dirac operator\footnote{$\Moo$, $\Mee$ correspond to
$1+T_{\rm oo}$ and $1+T_{\rm ee}$ respectively in~\cite{jliu,DellaMorte:2003jj}.},
we recall the Schur complements of the hermitian Dirac operator 
with respect to asymmetric 
and symmetric even-odd preconditioning $\hQa, \hQ$
\be
\hQa = \hat{c}\,\gf (\Moo-\Moe \Mmee \Meo)\,,\qquad \hQ=\Mmoo \hQa
      \,,\qquad \hat{c}=(1+64\kappa^2)^{-1}.
\ee
The action then reads
\be
S = S_G ~+~ S_{\rm pf} ~+~ S_{\det},
\ee
with
\ba
S_{\rm pf} &=&\phi_0\dg \left[ \hQ\hQ\dg+\rho_0^2\Moo^{-2}\right]^{-1} \phi_0 +
              \phi_1\dg \left[\rho_0^{-2}+
    \hQa^{-2} \right]\phi_1  \la{eq:S_pf} \\
S_{\det} &=&   (-2)\log\det \Mee ~+~ (-2)\log\det \Moo,  \la{eq:S_det}
\ea
and $S_G$ is the plaquette action.
The determinants appearing in $S_{\det} $ are taken into account exactly.

In \tab{tab:sim} and \tab{tab:algo} we list the simulations discussed in this paper.
The reference length scale $L^*$ is defined through $\bar g^2(L^*)=5.5$,
where $\bar g$ is the Schr\"odinger functional coupling,
and the values it assumes at the relevant bare couplings were presented 
in~\cite{lat07:rainer}. For an estimate of $L^*$ in fermis, 
one may use the result $a=0.0784(10)$fm at $\beta=5.3$~\cite{cernI}, 
yielding $L^*\approx 0.6$fm.


Renormalization is carried out non-perturbatively in the SF 
at the scale $\mu_{\rm ren}=1/L_{\rm ren}$, where $\bar g^2(L_{\rm ren})=4.61$.
The values of the renormalization factor $\zp$ of the 
pseudoscalar density are taken from~\cite{Della Morte:2005rd}, while the values of 
the renormalization factor $\za$ of the axial current differ
from~\cite{Della Morte:2005rd}. They are presently
re-evaluated using a Ward identity in a $1.8\,\fm \times (1.2\,\fm)^3$ 
Schr\"odinger functional where the O$(a^2)$
effects are significantly smaller than before.
In the table we list our preliminary numbers~\cite{za:prel},
which are not expected to change by more than the quoted errors. 

\begin{table}[htbp] \small
\begin{center}
\begin{tabular}{c@{~~~~}c@{~~~~}c@{~~~~}c|c@{~~~~}c@{~~~~}c}
\toprule
sim.      & $\beta$  & $(L/a)^3\times T/a$ &  $\kappa$ &  $L^*/a$ & $\za$ & 
$\zp$ \\
\midrule
$A_1$ & 5.5   & $32^3\times42$ &   0.13630 & 10.68(15) & 0.805(5) & 0.5008(70) \\
\midrule
$B_{1},B_{1}'$ &  & & 0.13550 &        &  &    \\
$B_2$ & 5.3   & $24^3\times32$ &   0.13590 & 7.82(6) &0.781(8)  & 0.4939(34)  \\
$B_3$ &      &                &   0.13605 &       &  &    \\
$B_4$ &       &                &   0.13625 &      &  &     \\
\midrule
$C_1$  & 5.2  & $16^3\times32$ &   0.13568 & 6.51(12) & 0.769(12) &  0.4788(5) \\
$C_2$  &      & $24^3\times32$ &   0.13568 &     &  &       \\
\bottomrule
\end{tabular}
\end{center}
\caption{Simulation parameters. We use $L^*$, defined by $\gbar^2(L^*)=5.5$,
  as a reference scale.
The renormalization factor of the axial current \cite{Della
  Morte:2005rd,za:prel}, $\za$, 
and of the pseudoscalar density ~\cite{DellaMorte:2005kg} 
at scale $\mu_{\rm ren}$ are listed.}
\la{tab:sim}
\end{table}

\begin{table}[htbp] \small
\begin{center}
\begin{tabular}{c@{~~}c@{~~}c@{~~}c@{~~~}c@{~~~}c@{~~~}c}
\toprule
  & mol. dyn.  & $N_{\rm rep}\cdot\tau_{\rm tot}$ & $\rho_0$ &
 $\<N^{(0)}_{\rm CG}\> $ & $\< N^{(1)}_{\rm CG}\>$  & $P_{\rm acc}$\\
\toprule
$A_1$ & [LF; 2; 5; 50] & $1 \cdot 4340 $ & 0.019803 & 170 & 824& 88\% 
\\
$B_{1}$ & [SW; 2; 1; 64] & $2\cdot 2400$&0.0300&100 & 482 & 91\%\\
$B_{1}'$ &[SW; \hbox{$\frac{1}{2}$}; 1; 16] &$2 \cdot 1750$& 0.0300 &  100 & 485 & 90\%\\ 
$B_2$ & [SW; \hbox{$\frac{1}{2}$}; 1; 16] & $2\cdot 1900$&0.0300&102 & 729 & 90\%\\
$B_3$ & [LF; 2; 5; 50] & $2\cdot 2600$ &0.019803 & 143&905&91\%\\
$B_4$ & [LF; 2; 5; 50] & $2\cdot 1448$ &0.0180 & 155&1195&87\%\\
$C_1$ & [LF; 2; 5; 64] &$1\cdot 6500$ & 0.0198 & 179 &791&96\%\\
$C_2$ & [LF; 2; 5; 80] &$2\cdot 2080$ & 0.0198 & 184 &1086&94\%\\
\bottomrule
\end{tabular}
\end{center}
\caption{Algorithmic parameters of the simulations. The molecular dynamics is 
characterized by $[{\rm Integrator};\tau;\delta\tau_1/\delta\tau_0;\tau/\delta\tau_1]$,
where the integrator can be `leap-frog' or `Sexton-Weingarten' and subscripts refer
to the two pseudofermions in use. For the gauge force, the SW integrator with 
$\delta\tau_{0}/\delta\tau_{\rm g}=4$ is used 
in all cases, and $\< N^{(k)}_{\rm CG}\>$ is the number of conjugate-gradient iterations
used to solve the symmetrically even-odd preconditioned Dirac equation during the 
trajectory.}
\la{tab:algo}
\end{table}

\subsection{Stability and the spectral gap}
The spectral gap $\mu$ of the Hermitian Dirac operator was used in~\cite{cern05}
as a tool to diagnose the stability of the HMC algorithm.
We define 
\be
\hat\mu={1 \over 4 \kappa \hat{c}}
        {\rm min}\left\{\sqrt{\lambda} ~{\Big|} \lambda \textrm{ is an eigenvalue of } \hat Q\hat Q^\dagger\right\}\,,
%
\ee
normalized such that it is given by the quark mass in the free theory 
with periodic boundary conditions. 
Since the only term that can potentially 
lead to unbounded fluctuations of the molecular dynamics forces
is associated with $\hat Q$, a sufficient condition for the 
stability of the algorithm is for the distribution of $\hat\mu$ 
to be well separated from the origin. 
We remark that 
$\hat\mu$ and $\mu$ (which was considered in~\cite{cern05})
cannot be directly compared on a quantitative level as they differ
by the boundary conditions in the time direction and due to our
(symmetric) even-odd preconditioning.
We obtained $\hat\mu$ by computing the lowest eigenvalue
 of $\hat Q\hat Q^\dagger$ using the 
algorithm of~\cite{Kalkreuter:1995mm}. 
Figure \ref{fig:lhsqrt} displays the histogram of $\hat\mu$
for simulations $C_{1,2}$. 
There is a clear separation of the median of the distribution 
from the origin, but in a few cases in the course of the simulations
eigenvalues as small as a third of this value were seen. 

We consider now the variance $\hat\sigma^2$ of $\hat\mu$. In~\cite{cern05}, 
a measure $\sigma$ of the width of the $\mu$ distribution was found to approximately 
satisfy  $a\sigma\sqrt{L^3T/a^4} \approx {\rm constant}$.
In the subset of our simulations where we computed $\hat\mu$, we find 
\be
\hat\sigma\sqrt{L^3T}/a=\left\{\begin{array}{l@{\quad}l}
            1.437(64)  & A_1 \\ 
            1.268(23) & C_1 \\
            1.477(33)  & C_2\,,
                   \end{array}\right.
\ee
varying only by
about $15\%$.

\begin{figure}
\begin{center}
\psfig{file=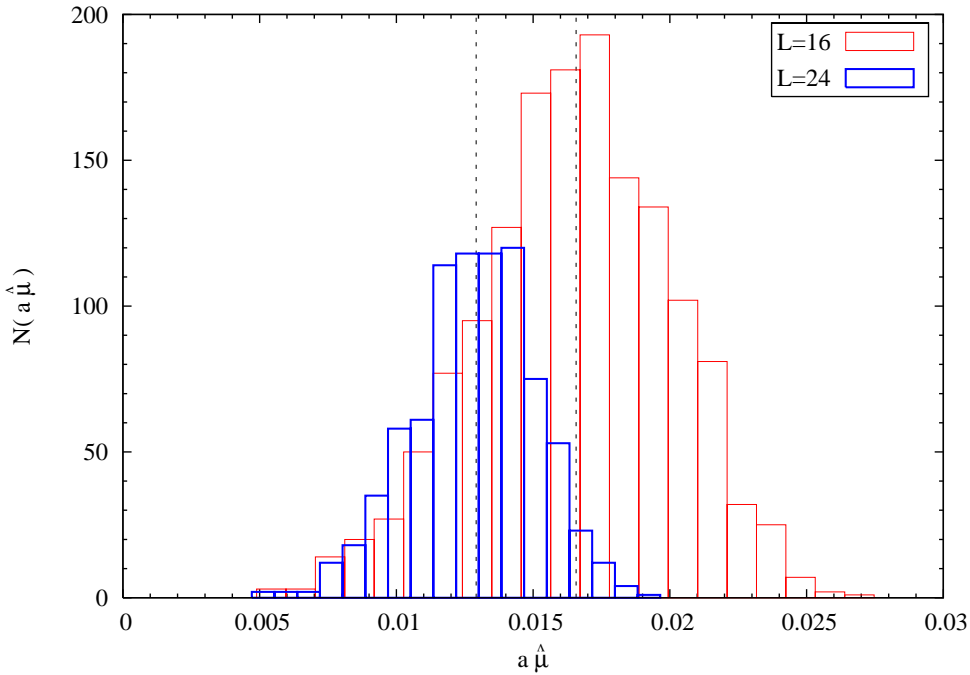,angle=-90,width=13cm}
\end{center}
\caption{Histogram of $\hat\mu$
for two different spatial volumes, simulations $C_1$ and $C_2$.
The median is indicated in each case by the vertical dashed line.}
\la{fig:lhsqrt}
\end{figure}

\subsection{Autocorrelation times}
We compile observed integrated autocorrelation times
$\tauint$  \cite{uwerr} in \tab{tab:perf} for five quantities discussed
and defined in detail in the next section.
The dependence of the autocorrelation times on the trajectory 
length was discussed previously~\cite{tlength}.
Here we note that while there is a tendency for the 
autocorrelation time of the plaquette to decrease when the 
lattice spacing is increased, for other observables the
opposite trend appears to be present underlining that 
autocorrelations have to be monitored for each observable 
separately. The most important information in \tab{tab:perf} is
that all autocorrelations are small compared to the length
of the runs (cf. \tab{tab:algo}). Error estimates are hence trustworthy.

\begin{table} [htbp]\small
\begin{center}
\begin{tabular}{c@{~~~}c@{~~~}c@{~~~}c@{~~~}c@{~~~}c@{~~~}c@{~~~}c}
\toprule
$\tauint[O]$  & $P$ & $m(T/2)$ & $\meffa(T/2)$& $\meffp(T/2)$ & $\feff(T/2)$ & $\meffv(T/2)$&$\geff(T/2)$\\
\midrule
$A_1$ &    5.0(9) & 4.9(9)& 11(3) & 21(6) & 10(2) & 40(10)&23(7) \\
$B_{1}$ & 13(3) & 5.5(9)& 7(1) & 16(4) & 4.2(7) & 23(7)&11(3) \\
$B_{1}'$ & 6(1) & 6(1)& 10(2) &22(7) & 14(4) & 24(8)&12(3)\\
$B_2 $ & 4.1(7) & 4.1(7) & 10(3) & 14(4) & 8(2) & 23(7)&24(8) \\
$B_3 $ & 9(2) & 3.9(6) & 4.7(7) & 11(2) &6(1) & 11(3) &11(2)\\
$B_4 $ & 8(2) & 5(1) & 6(1) & 7(2) &4.6(9) & 15(5) &8(2)\\
$C_1 $ & 9(2) & 5.3(8) & 5.2(8) & 5.1(8) &4.7(7) & 4.9(7)&5.6(9) \\
$C_2 $ & 11(3) & 6(1) & 6(1) & 7(1) & 3.9(6) & 6(1)&6(1) \\
\bottomrule
\end{tabular}
\caption{The integrated autocorrelation times for the plaquette, the current quark mass, 
the effective pseudoscalar mass and decay constant, and the effective vector mass. 
The unit is molecular dynamics time, i.~e. trajectories times the length of
the trajectory. For a precise definition of the observables see the following section.}
\label{tab:perf}
\end{center}
\end{table}

\section{Scaling test}
In this section, which represents the central part of this paper, 
we investigate the cutoff effects on a number of non-perturbatively renormalized
quantities. In order to keep systematic effects due to a varying volume negligible,
we compare series of simulations in a fixed (but quite large) volume on a physical scale.
More precisely we determine
$L/L^*=3.00(4),\,3.07(3)$ and   $T/L^*=3.93(4),\,4.09(3)$ on the $A$ 
and $B$ lattices. At $\beta=5.2$,  the volumes came out less uniform,
$L(C_1)/L^*=2.46(5),\,L(C_2)/L^*=3.69(6)$ and
$T(C_i)/L^*=4.92(10)$. We shall discuss how to correct for
these small mismatches after introducing the 
finite volume observables of this study.

\begin{table}[t]
\begin{center}
\begin{tabular}{ccccccc}
\toprule
sim.  &   $a\,m$    & $a\,\meffa$ & $a\meffp$ & $a\,\meffv$ &  
  ${a\,\feff\over Z_A\,(1+\bA a\mq)}$ & ${a^2\,\geff\over Z_P\,(1+\bP a\mq)}$ \\
\toprule
$A_1$ & 0.015519(37) & 0.1800(20) & 0.1793(15) & 0.2821(50) & 0.05999(42)&0.0629(10) \\
$B_1$ & 0.03388(12)& 0.3272(18) & 0.3236(16)&0.4520(35) & 0.09451(41) &0.1507(14) \\
$B_2 $ & 0.019599(95) & 0.2391(35) & 0.2406(19) & 0.3953(51) & 0.08442(68) & 0.1267(22)\\
$B_3 $ & 0.01460(11) & 0.2118(24) &0.2066(17)& 0.3647(35) &0.07714(60) & 0.1170(13)\\
$B_4 $ & 0.00727(14) &  0.1423(55)& 0.1528(20) & 0.3058(69) &0.0698(11) & 0.0985(15)\\
$C_1 $ & 0.01401(21) &  0.2173(55) &0.2338(24)&  0.4354(60) & 0.0877(13) & 0.1637(25)\\
$C_2 $ & 0.01442(14) &  0.2328(39) &0.2261(15) & 0.4152(42) &0.08773(67) & 0.1614(15)\\
$C_{\rm I}$&  0.01431(19) &  0.2286(97) &  0.2282(63) &  0.410(14) &  0.08772(61) &  0.1620(17) \\
\toprule
\end{tabular}
\caption{Simulation results for the effective quantities evaluated at $x_0=T/2$.
The bare current quark mass has been averaged over $T/3 \leq x_0 \leq 2T/3$.
The last line gives the interpolation of $C_{1},~C_{2}$, including the 
corrections described in the text.}
\label{tab:effO}
\end{center}
\end{table}

They are extracted from the zero spatial momentum boundary-to-bulk 
correlation functions, $\fA(x_0)\,,\;\fP(x_0)$ in the pseudoscalar 
channel,  $\kV(x_0)$ in the vector channel and the boundary-to-boundary 
pseudoscalar correlator $f_1$ \cite{guagnelli}. We include the 
O$(a)$ improvement term proportional to $\ca$ \cite{Della Morte:2005se} 
in $\fAimpr = \fA +a \ca\partial_0\fP$. Effective masses and decay constants
\bes
   \meffa(x_0) &\equiv&  -\half(\partial_0^*+\partial_0) \log(\fAimpr(x_0)) \la{eq:meffa}
   \\
   \meffp(x_0) &\equiv&  -\half(\partial_0^*+\partial_0) \log(\fP(x_0)) \la{eq:meffp}
    \\
   \meffv(x_0) &\equiv&  -\half(\partial_0^*+\partial_0) \log(\kV(x_0)) \la{eq:meffv}
    \\
   \feff(x_0) &\equiv&  -2 \za{\fA(x_0)\,(1+\bA a\mq) \exp(\meffa(x_0)(x_0-T/2)) \over
                        \left(\fone\,\meffa(x_0)\, L^3 \right)^{1/2}}  \nonumber\\
              &=& -2 \za\,(1+\bA a\mq) {\fAimpr(T/2)  \over
                        \left(\fone\,\meffa(T/2)\, L^3 \right)^{1/2}} \quad
                        \mbox{at} \quad x_0=T/2 
   \\
   \geff(x_0) &\equiv&  2 \zp\,(1+\bP a\mq) 
              {\fP(x_0) \exp(\meffp(x_0)(x_0-T/2))\,\meffp(x_0)^{1/2} \over
                        \left(\fone\, L^3 \right)^{1/2}} \nonumber\\
              &=& 2 \zp\,(1+\bP a\mq){\fP(T/2) \,\meffp(T/2)^{1/2} \over
                        \left(\fone\,L^3 \right)^{1/2}} \quad
                        \mbox{at} \quad x_0=T/2 
\ees 
are related to ($L$-dependent) masses and 
matrix elements,
\bes
  \meffa(x_0) \approx \Mps \approx  \meffp(x_0)\,,\quad 
\meffv(x_0) \approx \Mv \,,\quad
   \feff(x_0) \approx \Fps \,,\quad \geff(x_0) \approx \Gps \,.
\ees
These relations hold in the limit
of large $x_0$ and $T$ up to correction terms \cite{guagnelli}
\bes
   O_{\rm eff}(x_0) &=& 
               O + \eta_O \exp(-(E_1-\Mps)\,x_0) + \tilde\eta_O
               \exp(-E_2\,(T-x_0))  + \ldots\,,
   \label{e:corr}
\ees 
where the coefficients $\eta_O$ and $\tilde\eta_O$ are ratios of matrix
elements, $E_1$ is the energy of the first excitation in the zero momentum
pion channel and $E_2$ in the vacuum channel. For not too small $L$ and
not too large $\Mps$ we expect $E_1\approx3\Mps$ 
and $E_2\approx2\Mps$. Our
results for the effective observables at $x_0=T/2$ are listed in
\tab{tab:effO}
together with the bare current quark
mass $m$ stabilized by averaging over $T/3 \leq x_0 \leq 2T/3$,
\bes
 m &=& {1\over n_2-n_1+1} \sum_{x_0/a=n_1}^{n_2}
m(x_0)\,, \quad n_1\geq T/3a\,,\; n_2\leq 2T/3a\\
 m(x_0)&=&\frac{\half(\partial_0^*+\partial_0)f_{\rm A}(x_0)
                     +\ca a\partial_0^*\partial_0 f_{\rm P}(x_0)}{2f_{\rm P}(x_0)}\,.
\ees

The results at $\beta=5.3$ can be compared directly to those of
\cite{cernII}, shown in \tab{tab:cernO}, for which the correction 
terms in \eq{e:corr}
can safely be neglected. In other words they correspond to 
$x_0,T\to \infty$. 
This allows us to estimate the effects
due to $T(C) > T(A)\approx T(B)$ in addition to those coming
from the mismatch in $L$.
\begin{itemize}
\item[1.]
For the matrix elements $\feff\,,\geff$ no systematic differences
between $B$ and $D$ lattices are visible. No correction due to 
$T$ is necessary.  We just interpolate the
$C_1$ and $C_2$ results in $L$ to $L/L^*=3$ using 
the Ansatz $a_1+a_2\,L^{-3/2}e^{-\Mps L}$, with $\Mps$ the pion mass on the larger
volume. A small systematic error is added linearly to the statistical one. 
It is estimated by comparing with the result from an alternative interpolation with
$a_1'+a_2'\,L^{-1}$.
\item[2.]
We observe $|\meffp(B)/\meffp(D) -1| \leq 0.03$ without a systematic
trend as a function of the quark mass. We take this into account
as a systematic error of 2\% on $\meffp(C)$ and\footnote{From 
\eq{e:corr} this finite $T$ effect scales with $\exp(-\Mps  T)$,
yielding a reduction of 3\% by a factor $[1-\exp(-\Mps L^*)]$ when
one considers the difference between $T\approx5L^*$ and the target $T=4L^*$.}
subsequently we interpolate in $L$ as
in 1. The numbers for $\meffa$ are not used further. 
\item[3.]
Finite $T$ effects are not negligible in the vector mass
($\meffv(B)/\meffv(D) -1 \approx -0.10\; \ldots\; -0.03$). We thus first
perform a correction for the finite $T$ effects using fits to 
\eq{e:corr} with $E_1 = 2(\Mps^2+(2\pi/L)^2)^{1/2}$, $E_2 = 2\Mps$. 
A systematic error of 50\% of this correction is included for the result. 
Next the finite $L$  correction is performed as above.
\end{itemize}

\begin{table}[t]
\begin{center}
\begin{tabular}{cccccc}
\toprule
sim.  &   $a\,m$    & $a\,\Mps$ & $a\Mv$ &  ${a\,\Fps\over Z_A\,(1+\bA a\mq)}$ 
      &  ${a^2\,\Gps\over Z_P\,(1+\bP a\mq)}$  \\
\toprule
$D_1$ & 0.03386(11)  & 0.3286(10) & 0.464(3) & 0.0949(13) & 0.1512(20)\\
$D_2$ & 0.01957(07)  & 0.2461(09) & 0.401(3) & 0.0815(10) & 0.1260(16)\\ 
$D_4$ & 0.00761(07)  & 0.1499(15) & 0.344(9) & 0.0689(13) & 0.1017(24) \\
\toprule
\end{tabular}
\caption{Observables from fits of \cite{cernII} i.e. $x_0,T\to \infty$.
Input parameters $\beta$,
$\kappa$ and $L/a$ match those of lattices $B_1,B_2,B_4$; note that 
$D_4$ has been renamed here compared to \cite{cernII}.
}
\label{tab:cernO}
\end{center}
\end{table}

\begin{figure}[p]
\begin{center}
\psfig{file=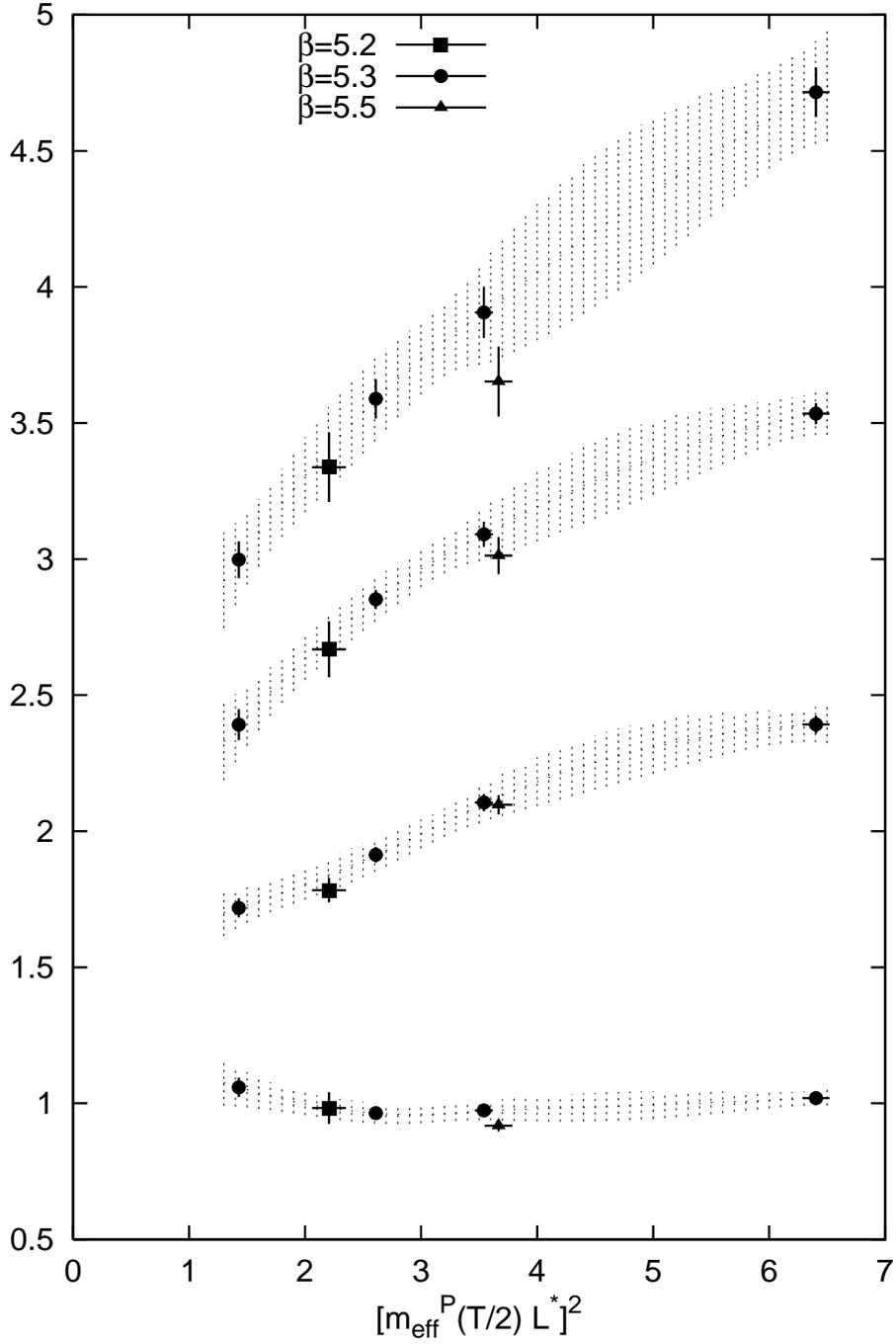,angle=0,width=13cm}
\end{center}
\caption{Dimensionless renormalized finite volume observables
as a function of $[\meffp L^*]^2$. From top to
bottom $\geff (L^*)^2,\;\meffv L^*,\; 4\feff L^*\,,\; [\meffp L^*]^2 /
 [\mbar(\mu_{\rm ren})L^*]/15$
 are shown. Squares, circles and triangle
are for $\beta=5.2\,,\;5.3\,,\;5.5$ respectively. 
Effective quantities are at $x_0=T/2$.
The dotted band is an interpolation of the $\beta=5.3$ data as described in
the text.  
}
\la{f:scaling}
\end{figure}

\begin{figure}
\begin{center}
\psfig{file=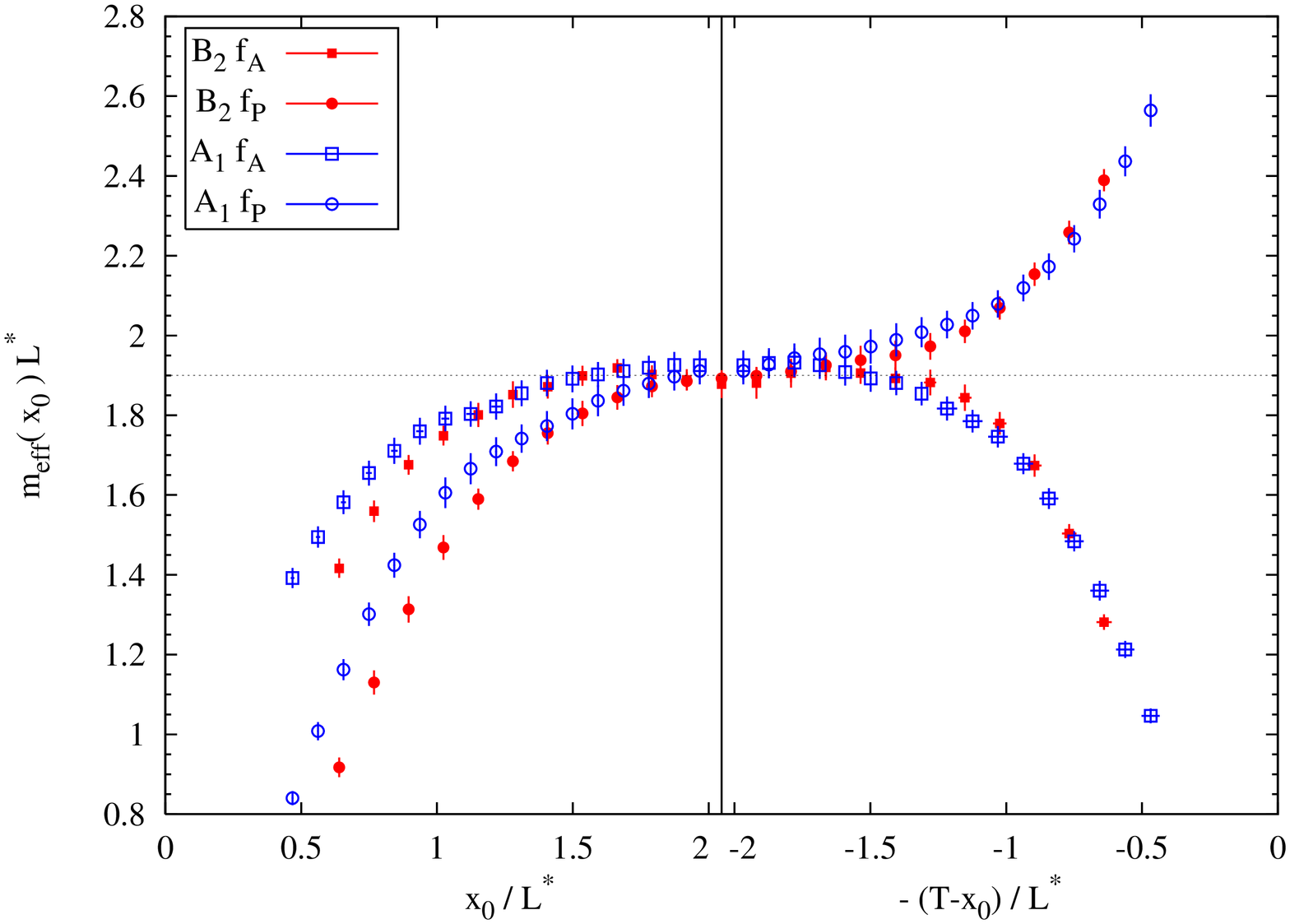,angle=-90,width=12.5cm}
\end{center}
\caption{The effective pseudoscalar masses $\meffa$ and $\meffp$
in simulations $B_2$ and $A_1$. 
The horizontal error bars are shown on some of the points only for clarity. 
The horizontal line is to guide the eye. The vertical line indicates the middle
of the $B_2$ lattice.}
\la{f:Mpseff}
\end{figure}

The interpolated values are
included in \tab{tab:effO} as ``simulation'' $C_{\rm I}$.
After these small corrections we are ready to look at the lattice 
spacing dependence of our observables.
To this end the necessary renormalization factors are attached (with
perturbative values for $\bA,\bP$ \cite{impr:pert}) and we form
dimensionless combinations by multiplying with $L^*$. 
At lowest order in the quark mass expansion (in large volume), one has
$M_{\rm PS}^2 \propto m$. It is thus natural to consider 
$[\meffp L^*]^2 / [\mbar(\mu_{\rm ren})L^*]$ instead of the quark
mass itself.  
We choose $\mbar$ renormalized non-perturbatively in the SF scheme at 
scale $\mu_{\rm ren}=1/L_{\rm ren}$ where $\gbar^2(L_{\rm ren}) =4.61$
\cite{DellaMorte:2005kg}. The quantities considered are shown in
\fig{\ref{f:scaling}} as a function of the dimensionless $[\meffp L^*]^2$. 
At $\beta=5.3$ we have a few quark-mass points. 
As a reference, these are locally interpolated in 
$[\meffp L^*]^2$ with a second order polynomial.
For masses lighter than in simulation $B_2$, 
the interpolation involves the lightest three masses and 
for heavier ones, it involves the heaviest three masses.
The two-sigma bands ($\pm 2 \sigma$) of these interpolations 
are depicted as dotted vertical lines. Our results at the other
$\beta$-values are seen to be in agreement with these error bands,
which are generally around 5\%, 
but 10\% for $[\meffp L^*]^2 / [\mbar(\mu_{\rm ren})L^*]$
after all errors are included. 
Even if the precision is not very impressive, large
cutoff effects are clearly absent.

So far we have discussed the scaling of the 
ground state properties for a given symmetry channel.
We now turn to the size of cutoff effects 
affecting \emph{excited state} contributions to the correlators.
Figure \ref{f:Mpseff} compares the effective pseudoscalar masses $\meffa$
and $\meffp$  in simulation $A_1$ and $B_2$. 
The large size of the excited state contributions~\cite{lat07:rainer},
while a drawback in extracting ground state properties,
means that these functions are rather sensitive to the aforementioned 
cutoff effects.
Because the $A_1$ time extent is shorter by $4(1)\%$, on this figure
we have separately aligned the two boundaries of lattice $A_1$ and $B_2$.
We observe that the two data sets are consistent within uncertainties 
well before the function flattens off. 
With the exception of $\meffp$ for $x_0<T/2$,  the agreement sets
in at a distance to the closest boundary of about $L^*$, where it
is easily seen that several excited states contribute significantly
to the correlation functions. Altogether this figure
is evidence that the masses and matrix elements of the first excited state
in both the pion and vacuum channel have
scaling violations not exceeding the few percent level. But higher states
can have rather significant discretization errors.


\section{Conclusion and an updated value of $\Lmsbartwo$}

We carried out a finite size scaling test of the standard non-perturbatively
O$(a)$-improved \cite{impr:sw,impr:pap1,csw} Wilson theory with two flavors 
of dynamical fermions. In contrast to previous indications
\cite{lat03:rainer}, cutoff effects are rather small in the
present situation where the linear extent of the volume is around 1.6\,\fm .
In fact within our precision of about 5\% (collecting all errors) 
for effective masses and matrix elements, no $a^2$ effects are visible. 
Continuum extrapolations of data from (say) $0.08\,\fm \leq a \leq 0.04\,\fm$ 
lattices which can nowadays be simulated \cite{DDHMC1,DDHMC2}, 
seem very promising. Such a programme has been initiated~\cite{CLS}.
A complementary effort~\cite{Boucaud:2007uk} uses
the  twisted mass regularization of QCD~\cite{Frezzotti:2000nk}. Also in this
case linear $a$-effects are absent \cite{Frezzotti:2003ni} and the
O$(a^2)$ effects appear to be moderate \cite{Dimopoulos:2007qy}.

Finally we exploit the increased confidence
in the scaling behavior of the simulated lattice theory
to slightly refine our earlier estimate of the $\Lambda$-parameter.
In \cite{alpha2} the product $L^*\,\Lmsbar= 0.801(56)$
was computed non-perturbatively in the two-flavor theory.
Setting the scale through $r_0=0.5\,\fm$  
the value
$\Lmsbartwo = 245(16)(16)\,\rm MeV$ was obtained emphasizing
that more physical observables should be used in the future to set
the scale. Given the quality of scaling observed in the previous
section, it seems safe to assume that $L^* m_{K}$ in the continuum
limit differs by no more than 5\% from its value  
at  $\beta=5.3$ where 
$m_{\rm K}a = 0.197(10)$  from \cite{cernI,cernII} and $L^*/a=7.82(6)$ 
\cite{lat07:rainer} are known.\footnote{
We have used $m_{\rm K} = m_{\rm K,ref }$ 
 with an error of 5\%  where $m_{\rm K,ref }$ is defined in
\cite{cernI}.}
We then 
obtain $\Lmsbartwo/m_{\rm K} = 0.52(6)$ or $\Lmsbartwo = 257(26)\,\rm MeV$, 
where a 5\% uncertainty for a possible
scaling violation has been added to the error (in quadrature). 
The new estimate is a bit higher than the previous one \cite{alpha2}.

\section*{Acknowledgements}
We thank DESY/NIC for computing resources on the APE machines
and the computer team for support, in particular to run on the 
apeNEXT systems.
This project is part of ongoing algorithmic development within
the SFB Transregio 9 ``Computational Particle Physics'' programme.
We thank the authors of~\cite{cernI} for discussions and for communicating
simulation results prior to publication.


\end{document}